\documentclass[oneside,reqno]{amsproc}

\usepackage{amsmath}
\usepackage{epsf}
\usepackage{graphicx}

\begin{document}

\markboth{Homer G. Ellis}
{Explain Cosmic Acceleration?  First, Correct Einstein}

\title{EXPLAIN COSMIC ACCELERATION?  FIRST, CORRECT EINSTEIN\footnote
{Selected for honorable mention in the Gravity
Research Foundation 2006 Awards for Essays on Gravitation.}}

\author{HOMER G. ELLIS \\
   \it  Department of Mathematics, University of Colorado at Boulder \\
        Boulder, Colorado 80309, USA \\
        Homer.Ellis@Colorado.EDU}

\begin{abstract}
In creating his gravitational field equations Einstein unjustifiedly assumed
that inertial mass, and its energy equivalent, is a source of gravity.
Denying this assumption allows modifying the field equations to a form in
which a positive cosmological constant appears as a uniform density of
gravitationally repulsive matter.  This repulsive matter is identified as
the back sides of the `drainholes' (called by some `traversable wormholes')
introduced by the author in 1973, which attract on the high, front sides
and repel more strongly on the low, back sides.  The field equations with a
scalar field added produce cosmological models that `bounce' off a positive
minimum of the scale factor and accelerate throughout history.  The `dark
drainholes' that radiate nothing visible are hypothesized to constitute the
`dark matter' inferred from observation, their excess of negative active
mass over positive active mass driving the accelerating expansion.  For a
universe with spatial curvature zero, and the ratio of scale factor now to
scale factor at bounce equal to the Hubble radius over the Planck length,
the model gives an elapsed time since the bounce of two trillion years.  The
solutions for negative spatial curvature exhibit early stage inflation of
great magnitude in short times.  Cosmic voids, filaments, and walls are
attributed to separation of the back sides of the drainholes from the front,
driven by their mutual attractive--repulsive interactions.
\end{abstract}

\maketitle

\markboth{HOMER G. ELLIS}
{EXPLAIN COSMIC ACCELERATION? FIRST, CORRECT EINSTEIN}

\baselineskip=15pt

{\parindent 0.5in  \it{Keywords}: Cosmic acceleration; dark matter/energy;
inflation.}

\vskip 10pt

Cosmologists are perplexed by the discovery that the expansion of the
universe, long thought to be slowing down, is in fact speeding up.  This
acceleration seems to require, in addition to the mysterious, unseen `dark
matter' invoked to explain the assembling of visible matter into galaxies
and galactic clusters and superclusters, an even more mysterious `dark
energy' that acts in a gravitationally repulsive manner to cause the
acceleration.  The initial attempt to model this dark energy by Einstein's
cosmological constant $\Lambda$, attributing its source to a negative
pressure created in the vacuum by virtual particle pairs, runs up against a
discrepancy of at least fifty-five orders of magnitude between the tiny
$\Lambda$ required to explain the acceleration and the large $\Lambda$
predicted by the quantum mechanics of these virtual pairs.  A way out of
this conundrum is available, but it requires the correcting of an erroneous
assumption Einstein made in the early days of the general theory of
relativity, an assumption that has stood virtually unchallenged down
through the years.  The assumption in question is {\it not} the introduction
in 1917 of $\Lambda$ into the field equations of gravity, self-described by
Einstein as a mistake.  It had in fact already appeared in his 1916 paper
Die Grundlage der allgemeinen Relativit\"atstheorie~\cite{eins} setting out
the fundamentals of the general theory.

In its most transparent form the assumption is that inertial mass, and
concomitantly its energy equivalent, is a source of gravity and must
therefore be coupled to the gravitational potential in the field equations
of the general theory.  Einstein arrived at this assumption in the 1916
paper while seeking a tensorial equation to correspond to the Poisson
equation $\nabla^2 \phi = 4 \pi \kappa \mu$, where $\mu$ denotes the
``density of matter''.  Drawing on the special theory's identification of
``inert mass'' with ``energy, which finds its complete mathematical
expression in . . . the energy-tensor'', he concluded that ``we must
introduce a corresponding energy-tensor of matter
$\text{T}^\alpha_\sigma$''.  Further describing this energy-tensor as
``corresponding to the density $\mu$ in Poisson's equation'', he wrote down
his now hallowed field equations
{$R_{\alpha \beta} - \frac12 R g_{\alpha \beta} =
\frac{8 \pi \kappa}{c^2} T_{\alpha \beta}$}.  The unjustified step in this
argument is the confusing of `gravitating mass', which is the sole
contributor to the ``density of matter'' in Poisson's equation, with
``inert mass'', which is indeed equivalent to energy in the proportion
$m = E/c^2$.  That all bodies respond alike to a gravitational field
establishes the equivalence of `inertial' (inert) mass with `passive'
(gravitat{\it ed}\/) mass, but there is no corresponding link between
passive and `active' (gravitat{\it ing}\/) mass, thus no link between
inertial mass and active mass.  In Newton's theory of gravity there is such
a link, but it depends on his law of action and reaction, which for
gravitating bodies would require instantaneous action at a distance,
something that in a relativistic field theory such as Einstein's does not
exist.  It likely is this inference from Newton's theory that caused
Einstein to treat ``inert mass'' and ``density of matter'' as equivalent.

If, contrary to Einstein's assumption, inertial mass and its energy
equivalent is not a source of gravity, then in particular the kinetic energy
in the form of the pressure $p$ in a continuous distribution of gravitating
matter must not contribute to gravity, so Einstein's choice
$T^{\alpha \beta} = \mu u^\alpha u^\beta + (p/c^2)(u^\alpha u^\beta -
g^{\alpha \beta})$ must be modified.  The most elegant way to effect a
modification is to not think about $T_{\alpha \beta}$, but instead derive
field equations as the Euler--Lagrange equations of the action integral
\begin{equation}
\int (R - \textstyle\frac{8 \pi \kappa}{c^2} \mu) (-g)^{\frac12} \, d^4\!x,
\label{eqn1}
\end{equation}
where $\mu$ is the {\it active} gravitational mass density.  (This integral
is the most straightforward relativistic analog of the action integral that
yields the Poisson equation for the newtonian gravitational potential.)
Variation of the metric produces the modified field equations
\begin{equation}
R_{\alpha \beta} - \textstyle{\frac12} R \, g_{\alpha \beta} =
-\frac{4 \pi \kappa}{c^2} \mu \, g_{\alpha \beta},
\label{eqn2}
\end{equation}
which makes $T_{\alpha \beta} = -\frac12 \mu g_{\alpha \beta}$.  Equivalent
to Eq.~(2) is
$R_{\alpha \beta} = \frac{4 \pi \kappa}{c^2} \mu \, g_{\alpha \beta}$, the
00 component of which reduces in the slow motion, weak field approximation
precisely to the Poisson equation.

Incorporation of other putative determinants of the geometry of space-time,
such as scalar fields and electromagnetic fields, can be accomplished in the
usual way by adding terms to the action integrand.  In particular, a
cosmological constant term can be added, changing the integrand to
$R - \textstyle\frac{8 \pi \kappa}{c^2} \mu + 2 \Lambda$ and the field
equations to
\begin{equation}
R_{\alpha \beta} - \textstyle{\frac12} R \, g_{\alpha \beta} =
-\frac{4 \pi \kappa}{c^2} (\mu + \bar \mu) \, g_{\alpha \beta},
\label{3}
\end{equation}
where $\frac{4 \pi \kappa}{c^2} \bar \mu = -\Lambda$.  Seen in this light a
positive $\Lambda$ is simply a (mis)repre\-sentation of a negative active
mass density $\bar \mu$ of a continuous distribution of gravitationally
repulsive matter, an excess of which over the positive active mass density
$\mu$ of attractive matter could drive an accelerating cosmic expansion.
Here one sees a glimmer of a solution to the `Cosmological Constant
Problem'.  The question is: Why should such gravitationally repulsive
matter exist and where should we look for~it?

In 1973 I described in considerable detail a model of a gravitating particle
alternative to the Schwarzschild vacuum solution of Einstein's field
equations.  This space-time manifold, which I termed a `drainhole', has
subsequently come to be recognized as an early (perhaps the earliest)
example of what is now called by some a `traversable wormhole'
\cite{elli,clem,mtor}. The
metric is a static, spherically symmetric solution of the field equations
$R_{\alpha \beta} - \textstyle{\frac12} R \, g_{\alpha \beta} =
T_{\alpha \beta} := -2 (\phi_{.\alpha} \phi_{.\beta} -
\textstyle{\frac12} \phi^{.\gamma} \phi_{.\gamma} \, g_{\alpha \beta})$ and
$\square \phi := \phi^{.\gamma}\!{}_{:\gamma} = 0$
arising from the action integrand $R + 2 \phi^{.\gamma} \phi_{.\gamma}$.
(N.B.  $R_{\alpha \beta}$ and $R$ here are the negatives of those
in~\cite{elli}.)  It has the proper-time form (in units in which $c = 1$)
\begin{equation}
\aligned
d \tau^2 &= [1 - f^2(\rho)] \, dT^2 - [1 - f^2(\rho)]^{-1} \, d\rho^2
                                    - r^2(\rho) \, d\Omega^2 \\
         &= dt^2 - [d\rho - f(\rho) \, dt]^2 - r^2 (\rho) \, d\Omega^2,
\endaligned
\label{4}
\end{equation}
where $t = T - {\displaystyle \int} f(\rho) [1 - f^2(\rho)]^{-1} \, d\rho$,
\begin{equation}
r(\rho) = \sqrt{(\rho - m)^2 + a^2} \, e^{(m/n) \alpha(\rho)}
\quad \text{and} \quad
1 - f^2(\rho) = e^{-(2m/n) \alpha(\rho)},
\label{5}
\end{equation}

\begin{equation}
\phi = \alpha(\rho) = \frac{n}{a}
               \left[\frac{\pi}{2}
                     - \tan^{-1} \left(\frac{\rho - m}{a}\right)\right],
\label{6}
\end{equation}
and $a := \sqrt{n^2 - m^2}$, $m$ and $n$ being parameters satisfying
$0 \leq m < n$.  (The coordinate $\rho$ used here is the $\rho$
of \cite{elli} shifted upward by $m$.)  The shapes and asymptotics of $r$
and $f^2$ are shown in Fig. 1.  Not shown, but verifiable, is that
$f^2(\rho) \sim 2m/\rho$ as $\rho \to \infty$.

\vskip 10pt
\epsfxsize=6.0truein
\centerline{\epsfbox{fig1.epsi}}
\vskip 5pt
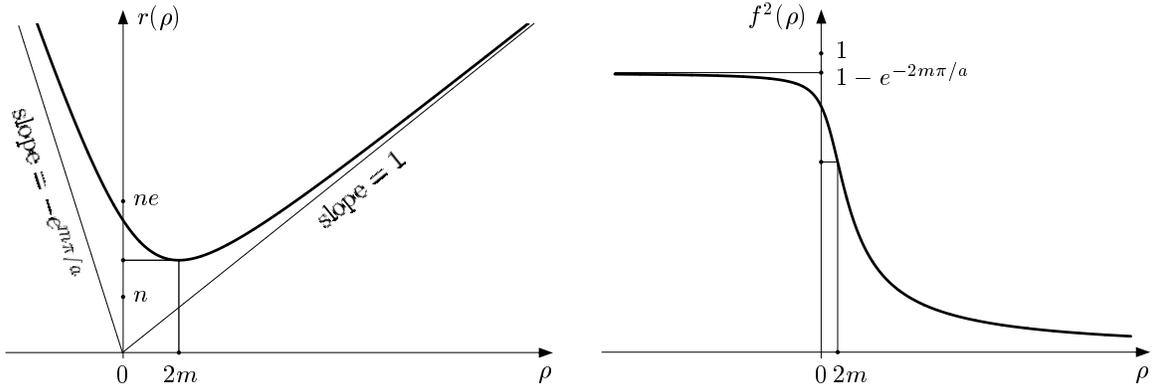
\centerline{Fig. 1.  Graphs of $r(\rho)$ and $f^2(\rho)$ for typical values
of the parameters $m$ and $n$.}
\vskip 10pt

The choke point of the drainhole throat is the 2-sphere at $\rho = 2m$, of
superficial radius $r(2m)$ (i.e., of surface area $4 \pi r^2(2m)$), which
increases monotonically from $n$ to $ne$ as $m$ increases from 0 to $n$.
Thus the size of the throat is determined almost exclusively by $n$,
independently of $m$.  Although the scalar field $\phi$ has a nonkinetic
`energy' density that contributes to the space-time curvature through
$T_{\alpha \beta}$, this energy has little to do with the strength of
gravity (as determined by $m$), rather is associated with the negative
spatial curvatures found in the
open throat, the negativity of which mandates the minus sign at the front of
$T_{\alpha \beta}$.  Because $r(\rho) \geq n > 0$ and $f^2(\rho) < 1$, the
space-time manifold is geodesically complete and has no one-way event
horizon, the throat being therefore traversable by test particles and light
in both directions.  The manifold is asymptotic as $\rho \to \infty$ to a
Schwarzschild manifold with (active gravitational) mass parameter $m$.  The
flowing `ether' (a figurative term for a cloud of inertial observers
free-falling geodesically from rest at $\rho = \infty$) has radial velocity 
$f(\rho)$ (taken as the negative square root of $f^2(\rho)$) and radial
acceleration $(f^2/2)'(\rho)$, which computes to $-m/r^2(\rho)$ and
therefore is strongest at $\rho = 2m$.  Because the radial acceleration is
everywhere less than~0, the drainhole attracts test particles on the high,
front side, where $\rho > 2m$, and repels them on the low, back side, where
$\rho < 2m$.  Moreover, the manifold is asymptotic as $\rho \to -\infty$ to
a Schwarzschild manifold with mass parameter $\bar m = -m e^{m \pi/a}$, so
the drainhole repels test particles more strongly on the low side than it
attracts them on the high side, in the ratio
$-\bar m/m = e^{m \pi/\sqrt{n^2 - m^2}}$.  The drainhole is a kind of
natural accelerator of the `gravitational ether', drawing it in on the high
side and expelling it more forcefully on the low side, much as a leaf blower
does~air.

The 1973 paper has in it the following sentence: ``A speculative
extrapolation from the asymmetry between $m$ and $\bar m$ is that the
universe expands because it contains more negative mass than positive, each
half-particle of positive mass $m$ being slightly overbalanced by a
half-particle of negative mass $\bar m$ such that $-\bar m > m$.''  To bring
this idea to bear on the cosmological conumdrum, let us study solutions of
field equations that incorporate a positive mass density $\mu$, a negative
mass density $\bar \mu$ such that $-\bar \mu > \mu$, and a scalar field
$\phi$ as above.  The action integrand
$R - \frac{8 \pi \kappa}{c^2} (\mu + \bar \mu) +
2 \phi^{.\gamma} \phi_{.\gamma}$ combines these elements and yields the
field equations
\begin{equation}
R_{\alpha \beta} - \textstyle{\frac12} R \, g_{\alpha \beta} =
T_{\alpha \beta} :=
-\frac{4 \pi \kappa}{c^2} (\mu + \bar \mu) \, g_{\alpha \beta} -
2 (\phi_{.\alpha} \phi_{.\beta} -
\textstyle{\frac12} \phi^{.\gamma} \phi_{.\gamma} \, g_{\alpha \beta})
\label{7}
\end{equation}
and $\square \phi := \phi^{.\gamma}\!{}_{:\gamma} = 0$.  For a
Robertson--Walker metric $c^2 dt^2 - R^2(t) ds^2$ and a scalar field
$\phi = \beta(t)$ these reduce to
\begin{align}
\label{8}
3 \frac{\dot R^2/c^2 + k}{R^2}
 &= -\frac{4 \pi \kappa}{c^2} (\mu + \bar \mu) - \frac{{\dot \beta}^2}{c^2}, \\
\intertext{}
\label{9}
\frac{2}{c^2} \frac{\ddot R}{R} + \frac{\dot R^2/c^2 + k}{R^2}
 &= -\frac{4 \pi \kappa}{c^2} (\mu + \bar \mu) + \frac{{\dot \beta}^2}{c^2},
\end{align}
\vskip 6pt
\noindent and
\vskip 6pt
\begin{equation}
\square \phi
 = \frac{1}{c^2}\left(\ddot \beta + 3 {\dot \beta} \frac{\dot R}{R}\right)
 = 0,
\label{10}
\end{equation}
where $k$ = 1, 0, or $-1$, the uniform curvature of the spatial metric
$ds^2$.  In addition there is, corresponding to the identity
$T_{\alpha}\!{}^{\beta}\!{}_{:\beta} = 0$, the equation
$\frac{4 \pi \kappa}{c^2} d(\mu + \bar \mu) = -2 (\square \phi) d\phi = 0$,
which implies that the `accelerant' $A$, defined by
$A := -\frac{4 \pi \kappa}{c^2} (\mu + \bar \mu)$, is a constant, positive
under the assumption that $-\bar \mu > \mu$.  Equation~(10) integrates to
${\dot \beta}^2 R^6 = B c^2$, where $B$ also is a positive constant if
$\dot \beta \neq 0$.  Equations~(8)~and~(9) then are together equivalent
to
\begin{align}
\label{11}
\frac{1}{c^2} \frac{{\dot R}^2}{R^2}
 &= -\frac{4 \pi \kappa}{3 c^2} (\mu + \bar \mu) - \frac{k}{R^2}
                                          - \frac{{\dot \beta}^2}{3 c^2}
  = \frac{A}{3} - \frac{k}{R^2} - \frac{B}{3 R^6}
  = \frac{A R^6 - 3 k R^4 - B}{3 R^6} \\
\intertext{and}
\label{12}
\frac{1}{c^2} \frac{\ddot R}{R} &= -\frac{4 \pi \kappa}{3 c^2} (\mu + \bar \mu)
                                    + \frac{{2 \dot \beta}^2}{3 c^2}
                                 = \frac{A}{3} + \frac{2 B}{3 R^6}
                                 = \frac{A R^6 + 2 B}{3 R^6}.
\end{align}
\vskip 5pt

Four implications of these equations are immediate.  First, the scale factor
$R$ has a positive minimum value $R_{\text{min}}$ (namely, the only positive
root of the polynomial $A R^6 - 3 k R^4 - B$), which rules out a `big bang'
singularity, putting in its stead a `bounce' off a state of maximum
compression at time $t = 0$, when $R(t) = R_{\text{min}}$.  Also,
$R(t) \to \infty$ as $t \to \pm \infty$.  Second, $\ddot R$ is always
positive, so the universal expansion is accelerating at all times after the
bounce, and the universal contraction is decelerating at all times before
the bounce.  Third, the `Hubble parameter' $H$ (:=~$\dot R/R$) behaves
asymptotically as follows:
\begin{equation}
\frac{1}{c^2} H^2 = \frac{A}{3} - \frac{3 k R^4 + B}{3 R^6}
                     \to \frac{A}{3} \quad
                      \begin{cases} \! \text{from below if } k \geq 0 \\
                                    \! \text{from above if } k < 0
                      \end{cases}
                       \hskip -14pt \Bigg\} \text{ as } R \to \infty.
\label{13}
\end{equation}
Fourth, the `acceleration parameter' $Q$ (:= $(\ddot R/R)/(\dot R/R)^2$)
behaves this way:
\begin{equation}
                Q = 1 + \frac{k R^4 + B}{H^2 R^6}
                     \to 1 \quad
                     \begin{cases} \! \text{from above if } k \geq 0 \\
                                   \! \text{from below if } k < 0
                     \end{cases}
                      \hskip -14pt \Bigg\} \text{ as } R \to \infty.
\label{14}
\end{equation}

Lacking singularities and horizons, and being geodesically complete,
mathematical drainholes are more pleasing to the aesthetic sense than are
mathematical blackholes.  Because in principle they are able to reproduce
all the externally discernible aspects of physical blackholes that
mathematical blackholes reproduce, they are at least as satisfactory as the
latter for modeling centers of gravitational attraction.  In this role they
are more aptly called `darkholes', inasmuch as they can capture photons that
venture too close, but, unlike blackholes, must eventually release them,
either back to the attractive high side whence they came or down the throat
and out into the repulsive low side.  Thus one can imagine that at galactic
centers will be found not supermassive blackholes, but supermassive
darkholes instead.  This, however, is not the end of the story.  A central
tenet of the general theory of relativity is that every elementary object
that `has gravity' is a manifestation of a local departure of the geometry
of space-time from flatness.  If such an object has other properties
ascribed to it by quantum mechanics or quantum field theory, these must be
additional to the underlying geometrical structure.  I therefore propose the
hypothesis that every such elementary gravitating object is at its core an
actual physical drainhole/darkhole (a `dark drainhole' I will call it) ---
these objects to include not only elementary constituents of visible matter
such as protons and neutrons, or, more likely, quarks, but also the unseen
particles of dark matter whose existence is at present only inferential.  It
then becomes a question of to what extent this hypothesis, coupled with the
cosmological model described above, fits the current state of observational
cosmology.

The significant parameters of the model are $k$, $A$, $B$, $H$, $Q$, $t_0$
(the present epoch), and $R_{\text{min}}$ ($= R(0)$) and $R(t_0)$ (or
perhaps only $R(t_0)/R_{\text{min}}$).  Although the current suspicion that
space is perfectly flat ($k = 0$) is perhaps not applicable in the context
of this model, let us proceed for the moment on the presumption that it is.
It then becomes straightforward to integrate Eqs.~(11) and~(12), the result
being
\begin{equation}
R^3(t) = R_{\text{min}}^3 \cosh (\sqrt{3 A} \, c \, t),
\label{15}
\end{equation}
where $R_{\text{min}} = (B/A)^{1/6}$, from which follow
\begin{align}
\label{16}
H(t) &= c \, \sqrt{\frac{A}{3}} \tanh (\sqrt{3 A} \, c \, t)
      = ({\text{sgn }} t) \, c \,
        \sqrt{\frac{A}{3}
              \left(1 - \left[\frac{R_{\text{min}}}
                                   {R(t)}\right]^6\right)} \, , \\
\label{17}
 Q(t) &= 1 + \frac{3}{\sinh^2 (\sqrt{3 A} \, c \, t)}
       = 1 + \frac{3}{{[R(t)/R_{\text{min}}]^6 - 1}} \, , \\
\intertext{and}
\label{18}
c^2 A &= H^2(t) [Q(t) + 2]
       = 3 H^2(t)\left[1 + \frac{1}{[R(t)/R_{\text{min}}]^6 - 1}\right].
\end{align}
Of the present values of these parameters the only one that is reasonably
well determined by observations is $H(t_0)$, which currently is estimated to
be about 72 (km/sec)/Mpc.  After $H(t_0)$ is input the others are
determined by the ratio $R(t_0)/R_{\text{min}}$.  Knowing neither the
numerator, the denominator, nor the ratio itself, but suspecting that the
ratio is quite large, the best one can do is make guesses and calculate the
results.  In this spirit let us go to extreme limits and take $R(t_0)$~cm to
be the Hubble radius $c/H(t_0)$ (the `radius of the observable universe')
and $R_{\text{min}}$~cm to be the Planck length.  Then
$R(t_0)/R_{\text{min}} =
1.28 \times 10^{28} \, \text{cm}/ 1.62 \times 10^{-33} \text \, \text{cm} =
7.93 \times 10^{60}$, which makes $Q(t_0) = 1 + 10^{-365}$, $c^2 A =
1.63 \times 10^{-35}/\text{sec}^2 = 1.62 \times 10^{-20}/\text{yr}^2$, and
$t_0 = 1.91 \times 10^{12}$ years.  This value for $t_0$ encompasses 140 of
the $13.6 \times 10^9$ years predicted to have elapsed since the `big bang'
by the `standard' (or `concordance') model based on the
{Friedmann--Robertson--Walker} equations, an interval which in the present
instance would allow approximately only a doubling from $R_{\text{min}}$ to
$R(t)$.  Other guesses are left to the reader, but with the advice that so
long as $R(t_0)/R_{\text{min}} \gg 1$ the parameters other than $t_0$ will
differ little from those values calculated above.

Consider now the problem of estimating the ratio $-\bar \mu/\mu$.  From
$A = -\frac{4 \pi \kappa}{c^2} (\mu + \bar \mu)$ follows
$-\bar \mu/\mu = 1 + c^2 A/4 \pi \kappa \mu$.  If we assume, for simplicity's
sake, that the dark drainholes whose active mass density is $\mu$ all have at
each epoch the same mass and size parameters $m$ and $n$, then
$-\bar \mu/\mu = -\bar m/m = e^{m \pi/a}$, so that
$m/\sqrt{n^2 - m^2} = m/a = \ln (-\bar \mu/\mu)/\pi =
\ln (1 + c^2 A/4 \pi \kappa \mu)/\pi$.  This implies that
\begin{equation}
\left(\frac{m}{n}\right)^2 =
    \frac{[\ln (1 + c^2 A/4 \pi \kappa \mu)]^2}
         {\pi^2 + [\ln (1 + c^2 A/4 \pi \kappa \mu)]^2}\,.
\label{19}
\end{equation}
Because it is only the combination $-\frac{4 \pi \kappa}{c^2} (\mu + \bar \mu)$
(the `accelerant' A) that remains constant, it is possible for the density
$\mu$ to change over time in some arbitrary fashion if $\bar \mu$ changes to
compensate.  Indeed $\mu$ (and $\bar \mu$) might not change at all, which
would indicate a `steady-state' universe with continuous creation of dark
drainholes sustaining the expansion.  In that case $\mu$ would be at all
epochs the density at the present epoch, which according to current best
estimates is the critical density $\mu_{\text{c}}$ of the FRW standard
model, namely, $\mu_{\text{c}} = 3 H^2(t_0)/8 \pi \kappa =
9.7 \times 10^{-30} \, \text{g/cm}^3$.  From Eq.~(18) one has
$c^2 A/4 \pi \kappa \mu = 3 H^2(t_0)
\left[1 + 1/([R(t_0)/R_{\text{min}}]^6 - 1)\right]/4 \pi \kappa \mu_{\text{c}}
= 2 \left[1 + 10^{-366}\right] = 2.0$.  Equation~(19) then yields
$m/n = 0.33 \, $.  If $n$ is the Planck length, then
$m = 0.33 \, \left(1.6 \times 10^{-33}\right) \, \text{cm}, =
7.2 \times 10^{-6} \, \text{grams}$ (= 0.33 Planck mass) in $c = \kappa = 1$
units.  This makes the dark drainhole particles gravitate ({\it not}
`weigh') much more than protons and neutrons.  To maintain the density
$\mu_{\text{c}}$ these particles would have to be created at a rate that
would keep on average about one in every $10^9$ cubic kilometers, which
would keep them on average about one thousand kilometers apart.  To decrease
the mass $m$ and thereby increase the number density would require taking
$n$ smaller than the Planck length.

Now consider the more orthodox supposition that the active gravitational
mass content of the universe is unchanging, so that $\mu$ decreases in
inverse proportion to the cube of the scale factor $R$: thus
\linebreak
$\mu = \mu_0 \, [R_{\text{min}}/R(t)]^3 = \mu\vert_{t = t_0} \, [R(t_0)/R(t)]^3
= \mu_{\text{c}} \, [R(t_0)/R(t)]^3$, where
$\mu_0 = \mu_{\text{c}} \, [R(t_0)/R_{\text{min}}]^3 =
4.9 \times 10^{153} \, \text{g/cm}^3$, the density at
the time of maximum compression.  Equation~(19) now reads
\begin{equation}
\left(\frac{m}{n}\right)^2 =
    \frac{\left[\ln (1 + (c^2 A/4 \pi \kappa \mu_{\text{c}}) \,
                [R(t)/R(t_0)]^3)\right]^2}
         {\pi^2 + \big[\ln (1 + (c^2 A/4 \pi \kappa \mu_{\text{c}}) \,
                       [R(t)/R(t_0)]^3)\big]^2}\,.
\label{20}
\end{equation}
The ratio of $m$ to $n$ increases monotonically as $t$ goes from 0 to
$\infty$.  When $t = 0$, $m/n = 1.3 \times 10^{-183}$.  When $t = t_0$,
$m/n = 0.33$ as in the steady-state case.  As $t \to \infty$, $m/n \to 1$
(the flow of the `gravitational ether' through the drainholes grows
asymptotically to the maximum rate that the drainholes can accommodate).  In
contrast to the steady-state version, which drives the accelerating
expansion by continually producing new drainholes of fixed size and mass,
this version drives it by continuously increasing the masses of a fixed
population of equal-sized drainholes.  The same effect could result from a
mixture of the two.  Also, in neither case is it written in stone that the
sizes must be uniform --- only the ratio of $m$ to $n$ is determinate.  In
particular, $n$ for particle constituents of atomic nuclei should perhaps
be of the order of $10^{-52} \, \text{cm}$ to account for the implication of
newtonian gravitational theory that their active masses must bear some
approximate numerical proportion to their inertial rest masses.

\vskip 10pt
\epsfxsize=6.0truein
\centerline{\epsfbox{fig2.epsi}}
\vskip 5pt
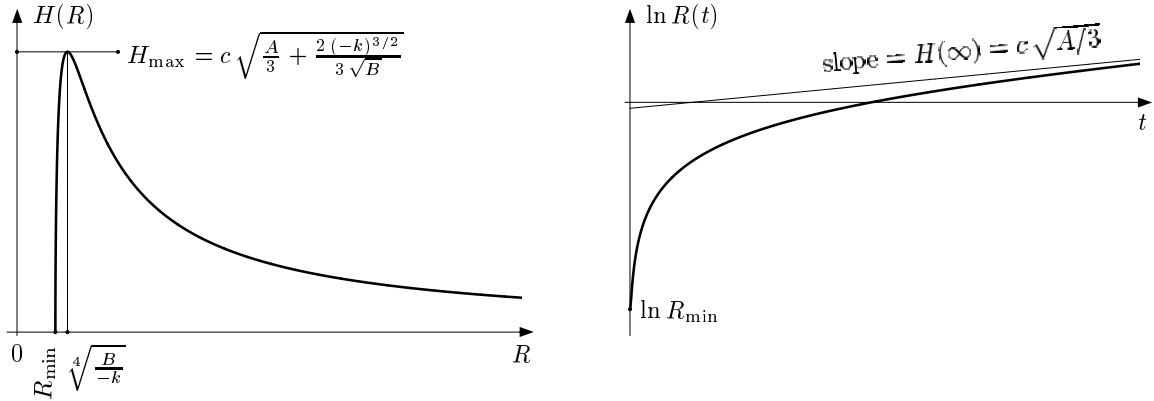
\centerline{Fig. 2.  Graphs of $H(R)$ and $\ln R(t)$, showing early stage
inflation for $k = -1$.}
\vskip 10pt

When $k = 0$ or 1 there is no early stage of extraordinary inflation, only a
steadily increasing Hubble parameter $H(t)$.  The situation is quite
different for $k = -1$ (strictly, $k = -1/\text{cm}^2$), as the graphs in
Fig. 2 demonstrate.  By sending $B$ toward~0 you can get as much inflation
as you might want --- and the more you get, the earlier you get it.  One can
show that $R_{\text{min}} \sim \root{4} \of {B/(-3k)}$ as $B \to 0$.  If, to
illustrate, $A = 9.09 \times 10^{-57}/\text{cm}^2$ and we take for
$R_{\text{min}}$ cm the Planck length $1.62 \times 10^{-33}$ cm,
then $B \approx 1.94 \times 10^{-131}/\text{cm}^2$, which gives $H(R)$ the
peak value $H_{\text{max}} =
c \, \sqrt{A/3 + 2 \, (-k)^{3/2}/(3 \, \sqrt{B})} = (5 \times 10^{60}) H(t_0) =
3.6 \times 10^{62} \text{(km/sec)/Mpc} = 1.17 \times 10^{43}/\text{sec}$,
occurring at $R = \root{4} \of {B/(-k)} =
2.10 \times 10^{-33} \approx 1.32 \, R_{\text{min}}$.  Numerical solution of
Eq.~(12) shows that at $t = 6.74 \times 10^{-14}$ seconds the scale factor
has inflated to $R(t) = 2.02 \times 10^{-3}$, for a ratio
$R(t)/R_{\text{min}} = 1.25 \times 10^{30}$, which works out to 100
doublings.  The acceleration $Q(R)$ is infinite at $R_{\text{min}}$, is
equal to 1 when $H(R) = H_{\text{max}}$, bottoms out with the value
$Q_{\text{min}} = 9.28 \times 10^{-82}$ at $R = 4.52 \times 10^{-13}$, when
$t = 1.51 \times 10^{-23}$ seconds, then returns slowly to 1 as
$R \to \infty$ (not until $R = 1.80 \times 10^{29}$ does $Q(R)$ reach 0.99,
at which time $t = 1.81 \times 10^{18}$ seconds $= 5.74 \times 10^{10}$
years).  Moving $B$ closer to 0 increases $H_{\text{max}}$ and drives
$R_{\text{min}}$ and $Q_{\text{min}}$ toward 0, and the 100-doublings time
toward 0 seconds.

A final question: If dark matter, which is understood to be distributed
unevenly in the universe, consists of the gravitationally attractive front
sides of dark drainholes, where do the repulsive back sides reside?  When
some local concentration of spatial curvature (a `quantum fluctuation', say)
develops into such a topological hole, the entrance and the exit, if close
together in the ambient space, will drift apart as the exit repels the
entrance more strongly that the entrance attracts the exit.  Apply this to a
multitude of particles and you will likely see the front side entrances
being brought together by both their mutual attractions and the excess
repulsion from the back side exits.  The exits, on the other hand, will tend
to spread themselves more or less uniformly over regions from which they
have expelled the entrances.  Herein lies a mechanism for producing the
voids, filaments, and walls of the cosmos.  What is more, the filaments and
walls should be more compacted than they would be if formed by gravitational
attraction alone, for the repulsive matter in the voids would increase the
compaction by pushing in on the clumps of attractive matter from many
directions with a nonkinetic, positive pressure {\it produced} by repulsive
gravity, not to be confused with the negative pseudo-pressure conjectured in
the confines of Einstein's assumption to be a {\it source} of repulsive
gravity.

It is tempting to speculate that Einstein might have taken some of the
steps described above had he recognized that his equating of active
gravitational mass with inertial mass was unjustified.  If there is a moral
here, it would be this one, which harks back to the discovery of
noneuclidean geometry that made possible the general theory of relativity:
Examine assumptions diligently, and when you find one you can't justify,
assume the opposite and see where it takes you.

\end{document}